\documentclass[aps,prl,reprint,superscriptaddress]{revtex4-1}

\usepackage{amssymb,amsmath,mathptmx}
\usepackage{graphicx}
\usepackage{siunitx}
\usepackage{floatrow}
\usepackage{relsize}
\usepackage{CJK}
\usepackage[breaklinks=true,colorlinks=true,bookmarks=false,urlcolor=blue,
citecolor=blue,linkcolor=blue]{hyperref}

\def\unit#1{\mathord{\thinspace\rm #1}}

\begin{document}

\begin{CJK*}{UTF8}{}
	
	\title{Tuning anti-Klein to Klein tunneling in bilayer graphene}
	\author{Renjun Du (\CJKfamily{gbsn}{杜人君})}
	\email{Renjun.Du@outlook.com}
	\affiliation{Institute of Nanotechnology, Karlsruhe Institute of Technology (KIT), D-76021 Karlsruhe, Germany}
	
	\author{Ming-Hao Liu (\CJKfamily{bsmi}{劉明豪})}
	\email{minghao.liu@phys.ncku.edu.tw}
	\affiliation{Institut f\"{u}r Theoretische Physik, Universit\"{a}t Regensburg, D-93040 Regensburg, Germany}
	
	\affiliation{Department of Physics, National Cheng Kung University, Tainan 70101, Taiwan}
	
	\author{Jens Mohrmann}
	\affiliation{Institute of Nanotechnology, Karlsruhe Institute of Technology (KIT), D-76021 Karlsruhe, Germany}
	
	\author{Fan Wu (\CJKfamily{gbsn}{吴凡})}
	\affiliation{Institute of Nanotechnology, Karlsruhe Institute of Technology (KIT), D-76021 Karlsruhe, Germany}
	\affiliation{College of Optoelectronic Science and Engineering, National University of Defense Technology, Changsha 410073, China}
	
	\author{Ralph Krupke}
	\affiliation{Institute of Nanotechnology, Karlsruhe Institute of Technology (KIT), D-76021 Karlsruhe, Germany}
	\affiliation{Institute of Material Science, Technische Universit\"{a}t Darmstadt, D-64287 Darmstadt, Germany}
	
	\author{Hilbert v. L\"{o}hneysen}
	\affiliation{Institute of Nanotechnology, Karlsruhe Institute of Technology (KIT), D-76021 Karlsruhe, Germany}
	\affiliation{Institute for Solid State Physics and Physics Institute, Karlsruhe Institute of Technology (KIT), D-76021 Karlsruhe, Germany}
	
	\author{Klaus Richter}
	\affiliation{Institut f\"{u}r Theoretische Physik, Universit\"{a}t Regensburg, D-93040 Regensburg, Germany}
	
	\author{Romain Danneau}
	\email{romain.danneau@kit.edu}
	\affiliation{Institute of Nanotechnology, Karlsruhe Institute of Technology (KIT), D-76021 Karlsruhe, Germany}
	
	\date{\today}

\begin{abstract}
We show that in gapped bilayer graphene, quasiparticle tunneling and the corresponding Berry phase can be controlled such that it exhibits features of single layer graphene such as Klein tunneling. The Berry phase is detected by a high-quality Fabry-P\'{e}rot interferometer based on bilayer graphene. By raising the Fermi energy of the charge carriers, we find that the Berry phase can be continuously tuned from $2\pi$ down to $0.68\pi$ in \textit{gapped} bilayer graphene, in contrast to the constant Berry phase of $2\pi$ in \textit{pristine} bilayer graphene. Particularly, we observe a Berry phase of $\pi$, the standard value for single layer graphene. As the Berry phase decreases, the corresponding transmission probability of charge carriers at normal incidence clearly demonstrates a transition from anti-Klein tunneling to nearly perfect Klein tunneling.
\end{abstract}

	\maketitle

\end{CJK*}

\paragraph{Introduction.} 

Bilayer graphene (BLG), like its single layer counterpart~\cite{novoselov2005two,geim2007rise,neto2009electronic,liu2011visualizing}, exhibits outstanding physical properties~\cite{morozov2008giant,novoselov2006unconventional,mccann2006asymmetry,mccann2013electronic} and is often regarded as promising materials for potential electronic applications. One striking feature of BLG is the possibility to induce and tune an electronic band gap by breaking the lattice inversion symmetry using, for example, an electric field~\cite{mccann2006asymmetry,oostinga2008gate,zhang2009direct,taychatanapat2010electronic,mccann2013electronic}. However, the fundamental knowledge of the gapped states in BLG remains limited in many respects despite the existing studies of the Berry phase~\cite{berry1984quantal,zhang2005experimental,novoselov2006unconventional,varlet2014fabry,varlet2015band} or quasiparticle tunneling~\cite{varlet2014fabry,katsnelson2006chiral,katsnelson2012graphene,kleptsyn2015chiral}.

\begin{figure}[b]
\includegraphics[width=8.5cm]{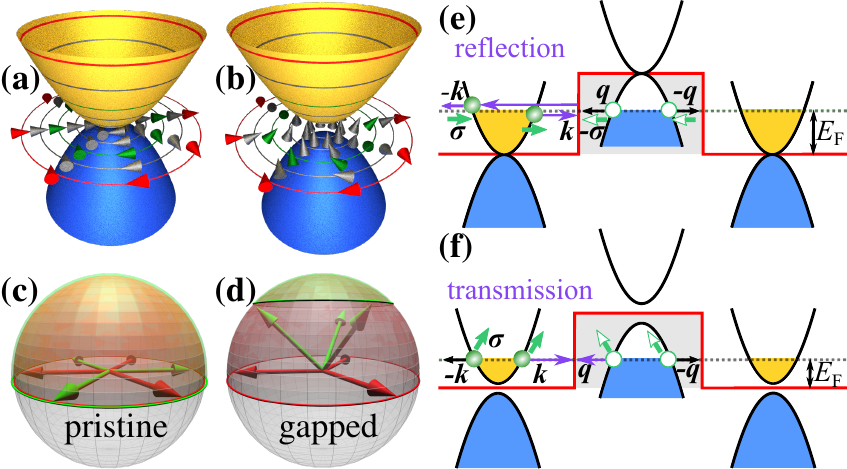}
\caption{Sketches of band structure and pseudospin orientation for pristine (a) and gapped (b) BLG. Pseudospin vectors at different Fermi levels (contours) are projected as small cones on a plane between the conduction (yellow) and valence (blue) bands. (c) and (d) show the corresponding Berry phase as the solid angle traced out by the pseudospin (arrows) on the Bloch sphere for (a) and (b), respectively. Red (green) color in (a)--(d) refers to high (low) Fermi energy ($E_\mathrm{F}$). (e) Anti-Klein tunneling for pristine BLG. $\boldsymbol{k}$ or $\boldsymbol{q}$ is the wave vector for electrons or holes. $\boldsymbol{\sigma}$ denotes the pseudospin. (f) Klein tunneling is possible in gapped BLG.}
\label{AKtoK}
\end{figure}

The emergence of a band gap has a strong impact on the Berry phase by modulating the pseudospin $\boldsymbol{\sigma}$~\cite{min2008pseudospin,macdonald2012pseudospin}, which expresses an extra quantum mechanical degree of freedom in graphene~\cite{novoselov2005two,katsnelson2006chiral}. In Figs.~\ref{AKtoK}(a)--(b) the pseudospin vectors at different Fermi levels are depicted as small cones and projected in a plane between the conduction (yellow) and valence (blue) bands in the momentum space. After a pseudospin vector adiabatically travels a closed path around the valley, \textit{e.g.}, the red circle in Figs.~\ref{AKtoK}(a)--(b), a Berry phase is acquired~\cite{xiao2010berry,varlet2015band,lundeberg2014harnessing,Ghahari845}. This process is better visualized on a Bloch sphere, as shown in Figs.~\ref{AKtoK}(c)--(d), where the pseudospin (denoted by arrows) traces out a solid angle which is equivalent to the Berry phase of BLG~\cite{xiao2010berry,varlet2015band}. In the absence of a band gap, \textit{e.g.}, in pristine BLG, the pseudospin vector always lies in the plane~\cite{macdonald2012pseudospin} (see Figs.~\ref{AKtoK}(a) and (c)), so the corresponding Berry phase remains $2\pi$~\cite{novoselov2006unconventional,mccann2013electronic} as shown by the half-spherical surface in Fig.~\ref{AKtoK}(c). On the other hand, the pseudospin may be polarized out of plane~\cite{min2008pseudospin,san2009pseudospin,macdonald2012pseudospin,lundeberg2014harnessing,varlet2015band} in gapped BLG (see Figs.~\ref{AKtoK}(b) and (d)), leading to a Berry phase in the range of 0--2$\pi$ as shown in Fig.~\ref{AKtoK}(d). The understanding of the tunable Berry phase in gapped BLG may shed light on the physical phenomena, such as the valley Hall effect~\cite{xiao2007valley,yao2008valley,gorbachev2014detecting,shimazaki2015generation}, the anomalous Hall effect~\cite{nagaosa2010anomalous,qiao2014quantum}, and quasiparticle tunneling~\cite{varlet2014fabry,varlet2015band}. A comprehensive exploration of the Berry phase in gapped BLG is, therefore, of fundamental interest.

The band gap also significantly affects quasiparticle tunneling, which is associated with the pseudospin~\cite{katsnelson2006chiral} and the Berry phase~\cite{varlet2014fabry,varlet2015band}. The quasiparticle tunneling in pristine BLG exhibits perfect reflection when the charge carriers encounter a sharp potential barrier at normal incidence, effect known as anti-Klein tunneling~\cite{katsnelson2006chiral,kleptsyn2015chiral}, as illustrated in Fig.~\ref{AKtoK}(e). However, when the band gap opens, anti-Klein tunneling can be reduced while the Berry phase slightly changes~\cite{varlet2014fabry}. 
Indeed perfect Klein tunneling, \textit{i.e.}, full transmission through a potential barrier~\cite{katsnelson2006chiral,katsnelson2012graphene,Shytov2008,Young2008,torres2014introduction}, may be possible in gapped BLG due to the out-of-plane polarization of the pseudospin~\cite{varlet2015band} (see Fig.~\ref{AKtoK}(f)). However, the observation of Klein tunneling in gapped BLG requires low-disorder devices and ballistic transport. To the best of our knowledge, such an anti-Klein to Klein tunneling transition has not been observed in BLG. 

In this paper, we employ an edge-connected hBN-BLG-hBN heterostructure (hBN for hexagonal boron nitride) to investigate quasiparticle tunneling in a lateral $pnp$ junction. We benefit from an advanced sample fabrication method~\cite{Wang2013a}, yielding ultra-clean devices, which enable ballistic Fabry-P\'{e}rot (FP) interferences~\cite{Rickhaus2013}. The phase-sensitive FP interference is used to detect the variation of the Berry phase. In contrast to previous work examining the Berry phase merely at high Fermi energies~\cite{varlet2014fabry}, the robust FP interference allows us to probe it close to the band edge. The role of the Berry phase and of the corresponding pseudospin on the quasiparticle tunneling will be discussed in detail and compared to numerical simulations based on a tight-binding model~\cite{Liu2012}.

\paragraph{Sample description.}

\begin{figure}
\includegraphics[width=85mm]{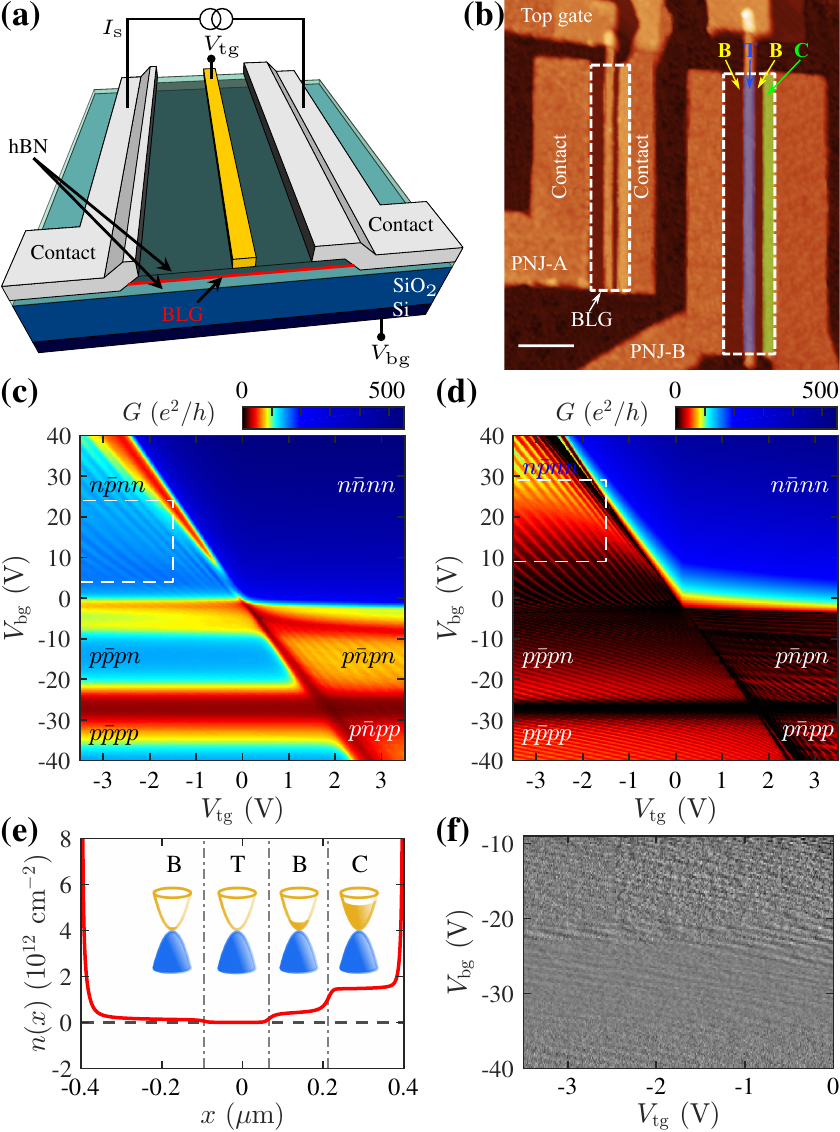}
\caption{Sketch (a) and AFM image (b) of the devices.  Scale bar in (b) is $1\unit{\si{\micro\meter}}$. Experimental (c) and simulation (d) results of conductance $G$ varying with $V_\mathrm{tg}$ and $V_\mathrm{bg}$ at $4.2\unit{K}$ and zero magnetic field for device PNJ-A. (e) The initial charge carrier density $n(x)$ across device PNJ-A when $V_\mathrm{bg}$ and $V_\mathrm{tg}$ are both zero. (f) Transconductance $dG/dV_\mathrm{tg}$ in the $p\bar{p}pn$ and $p\bar{p}pp$ regions of (c). }
\label{BL12D_FP_Gmap}
\end{figure}

The investigated devices, sketched in Fig.~\ref{BL12D_FP_Gmap}(a), consist of a hBN-BLG-hBN heterostructure. The encapsulation of BLG results in low-disorder devices, allowing ballistic transport over a distance of $9\unit{\si{\micro\meter}}$. The potential profile across the device is controlled by a local top gate about $150\unit{nm}$ wide as well as a global back gate (Si substrate). The fabrication follows Ref.~\onlinecite{Wang2013a}. Details of the devices are shown in Supplemental Material~\footnote{See Supplemental Material for full description of the sample information, electrostatic model, Fabry-P\'{e}rot interferences, the Berry phase and quasiparticle tunneling.}. Each device is divided into four regions, labeled as T (top- and back-gated region), B (only back-gated regions), and C (contact-overlapping region) in Fig.~\ref{BL12D_FP_Gmap}(b). The overlapping contact results in additional $n$-doping in region C when both gates are set to zero, as displayed in Fig.~\ref{BL12D_FP_Gmap}(e), where the carrier density profile is obtained from finite-element-based electrostatic simulation using \textsc{FEniCS} \cite{fenics} combined with the mesh generator \textsc{Gmsh} \cite{gmsh}.

\paragraph{Fabry-P\'{e}rot interferences.}

The conductance ($G$) as a function of the top- ($V_\mathrm{tg}$) and back-gate ($V_\mathrm{bg}$) voltages has been probed experimentally and modeled for device PNJ-A, as shown in Figs.~\ref{BL12D_FP_Gmap}(c) and (d), respectively. 
The conductance minima appear as three lines in these plots. The two horizontal lines at $V_\mathrm{bg}\approx-1.3\unit{V}$ and $V_\mathrm{bg}\approx-26.9\unit{V}$ are independent of $V_\mathrm{tg}$, and indicate the charge neutrality point (CNP) in regions B and C, respectively. The position of the CNP is determined by the initial doping of each region (see Fig.~\ref{BL12D_FP_Gmap}(e)). The diagonal line shows the CNP of the dual-gated region T and defines the displacement field axis, along which the interlayer asymmetry develops. The three lines partition the map into six sections, each of which has a unique combination of charge carrier polarities, as labeled on Figs.~\ref{BL12D_FP_Gmap}(c)--(d).

FP interferences arise in an electrostatic potential barrier with two semi-transmitting interfaces, if the phase difference $\Delta\Phi$ between two neighboring transmitted waves fits the resonance condition $\Delta\Phi=2\pi j$ ($j$ is integer). In the bipolar regime ($n\bar{p}nn$, $p\bar{n}pn$ and $p\bar{n}pp$), where the charge carrier type in region T (denoted by the overlined symbol) is different from adjacent region B, we observe clear conductance oscillations as a consequence of FP interferences. The FP fringes extend along the diagonal line, illustrating that the FP interference occurs in a cavity tuned by both $V_\mathrm{tg}$ and $V_\mathrm{bg}$. 
The cavity length is determined by the resonance condition of FP interferences as in Ref.\ \onlinecite{varlet2014fabry}, and is found to be around $150\unit{nm}$, which corresponds to the top-gate width. On the other hand, due to the long spacing between the contacts, FP interferences in unipolar regimes such as $p\bar{p}pn$ and $p\bar{p}pp$ are hardly visible. However, the weak oscillations become discernible in the transconductance $dG/dV_\mathrm{tg}$ gate map, see Fig.~\ref{BL12D_FP_Gmap}(f). 
More details about FP interferences are shown in Supplemental Material~\cite{Note1}.

To gain further insight into the implications and ramifications of our experimental results, quantum transport simulations based on the real-space Green's function method using the tight-binding model for Bernal-stacked BLG has been performed. Details of the simulation method are similar to Ref.\ \onlinecite{varlet2014fabry}, including how the gate-tunable interlayer asymmetry parameter $U$ can be implemented~\cite{mccann2013electronic}, with the following two alterations. First, the scalable tight-binding model \cite{liu2015scalable} with a scaling factor of $s_f=4$ has been adopted. Second, carrier density profiles obtained from electrostatic simulations [an example is shown in Fig.~\ref{BL12D_FP_Gmap}(e)] have been implemented in order to extract realistic on-site energy profiles for the tight-binding model Hamiltonian. More details about the gate-modulated carrier density profiles can be found in Supplemental Material~\cite{Note1}. Comparing Figs.~\ref{BL12D_FP_Gmap}(c) and (d), our experiment captures all the interference patterns that are theoretically predicted. This agreement demonstrates the high quality of both our FP interferometer design and the quantum transport simulations, even comparable to the suspended graphene interferometer with smooth junction profiles that led to high FP finesse~\cite{Rickhaus2013}. 

\paragraph{Berry phase and quasiparticle tunneling.}

\begin{figure}[t]
\includegraphics[width=8.5cm]{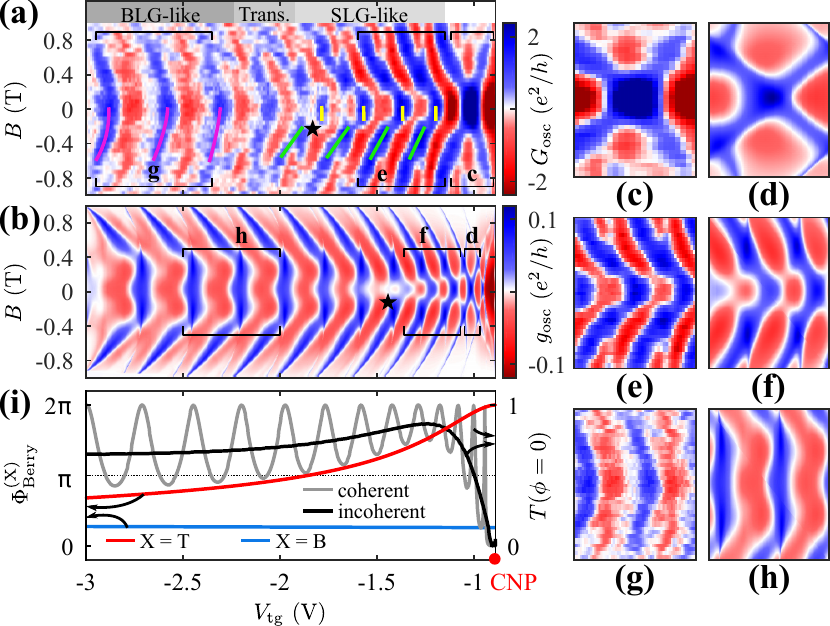}	
\caption{Fabry-P\'{e}rot interference measurements (a) and simulations (b) at $V_\mathrm{bg}=20\unit{V}$ under low magnetic fields for device PNJ-B (conductance measurements at zero magnetic filed are shown in Supplemental Material~\cite{Note1}).
The regions labeled by c--h in Figs. (a) and (b) are highlighted in the corresponding panels (c)--(h). (i) Berry phases for regions B and T are shown as blue and red curves, respectively. The corresponding transmission probability at normal incidence is calculated with phase-coherent (grey curve) and phase-incoherent (black curve) methods. }
\label{lowfield}
\end{figure}

At low magnetic fields, the phase difference $\Delta\Phi$ comprises not only the conventional kinetic part, the Wentzel-Kramers-Brillouin phase $\Phi_\mathrm{WKB}$, but also the Aharonov-Bohm phase $\Phi_\mathrm{AB}$ and the Berry phase $\Phi_\mathrm{Berry}$, which may arise under magnetic fields. The effect of the Berry phase on FP interferences may manifest itself as phase shifts of the FP fringes at certain magnetic fields~\cite{Shytov2008,Young2008,varlet2014fabry}, unlike $\Phi_\mathrm{WKB}$ and $\Phi_\mathrm{AB}$ yielding a continuous parabolic dispersion of the fringes with respect to $B$~\cite{varlet2015band}. Thus, the phase-sensitive FP interference is a convenient tool to probe the Berry phase generated in the cavity.

We observe the FP interference under low magnetic fields ($\left| B \right| \leq 0.9\unit{T}$) by tuning $V_\mathrm{tg}$ at fixed $V_\mathrm{bg}=20\unit{V}$, see  Fig.~\ref{lowfield}(a). Here, the oscillatory part of the conductance $G_\mathrm{osc}$ is presented instead of the total conductance $G$ in order to circumvent  the non-uniform conductance profile induced by the increasing magnetic field. We subtract a smoothed background $G_\mathrm{0}(V_\mathrm{tg})$ at each $B$ value, and obtain the oscillatory part via $G_\mathrm{osc}(V_\mathrm{tg})=G(V_\mathrm{tg})-G_\mathrm{0}(V_\mathrm{tg})$. The low-field dispersion of the FP fringes shows two distinct features. For $V_\mathrm{tg}$ close to the CNP ($V_\mathrm{tg}>-1.9\unit{V}$), the FP fringes shift suddenly from the initial positions (\textit{e.g.} yellow lines) to positions at slightly lower $V_\mathrm{tg}$ (\textit{e.g.} green lines) at $|B^{*}|\approx 0.15$--$0.25\unit{T}$, indicating that the Berry phase has been abruptly added to $\Delta\Phi$ ~\cite{Young2008,Shytov2008,RamezaniMasir2010fabry,liu2012efficient}.
The amount of phase shift for each fringe increases with decreasing $V_\mathrm{tg}$, and reaches $\pi$ at $V_\mathrm{tg}\approx-1.8\unit{V}$ (see the black star in Fig.~\ref{lowfield}(a)), suggesting that the Berry phase is continuously tuned across $\pi$ by modulating $V_\mathrm{tg}$. 
Here, the emergence of the Berry phase at $B^{*}$ instead of $B=0$, resembles the behavior of single-layer graphene (SLG), where the required geometric paths to acquire the Berry phase are formed with the assistance of low magnetic fields~\cite{Young2008,Shytov2008,RamezaniMasir2010fabry,liu2012efficient}.
For $V_\mathrm{tg}$ far away from the CNP ($V_\mathrm{tg}<-2.23\unit{V}$), the FP fringes exhibit parabolic dispersion (marked by magenta lines) with respect to $B$ as expected for BLG~\cite{varlet2014fabry}. This BLG-like dispersion without phase shift at $B^{*}$ illustrates that the Berry phase has already been included in $\Delta\Phi$ at $B=0$, highlighted by a transition region (labeled by Trans. in Fig.~\ref{lowfield}(a)) between the BLG-like and SLG-like dispersions. The reason is that the trajectory of the wave vector ($k$) forms a closed loop encircling the origin of momentum space, thus resulting in the non-zero Berry phase~\cite{varlet2015band}.

We have successfully reproduced the two types of dispersion, \textit{i.e.}, the SLG-like and the BLG-like, using quantum transport simulations based on a realistic electrostatic model, which is constructed from our experimental parameters but with a scattering region (the length $L=300\unit{nm}$) around the top gate, see Fig.~\ref{lowfield}(b). Note that $g_\mathrm{osc}$ is the oscillatory part of the calculated single-mode conductance $g$~\cite{Liu2012,Rickhaus2013,Note1}, and obtained using the same procedure as $G_\mathrm{osc}$. For better comparison, the fringes in the regions labeled by c--h are highlighted in Figs.~\ref{lowfield}(c)--(h). We found that the simulation result shows remarkable agreement with the experiment on the SLG-like (Figs.~\ref{lowfield}(e)--(f)) and BLG-like (Figs.~\ref{lowfield}(g)--(h)) dispersions under low magnetic fields, although the simulated patterns in Figs.~\ref{lowfield}(d), (f) and (h) occupy smaller regions in Fig.~\ref{lowfield}(b). In addition, the Berry phase of $\pi$ appears at $V_\mathrm{tg}\approx-1.45\unit{V}$ (black star) in Fig.~\ref{lowfield}(b) instead of near $-1.8\unit{V}$ due to the reasonable differences between the realistic electrostatic model and the intricate experiments.

We calculate the Berry phase by circular integral~\cite{varlet2014fabry,varlet2015band,xiao2010berry} for the gate range in Fig.~\ref{lowfield}(a) (see Fig.~\ref{lowfield}(i)). The Berry phase in region T (\smash{$\Phi^\mathrm{\!\mathsmaller{ (T)}}_\mathrm{\!\mathsmaller{Berry}}$}) is modulated from $2\pi$ to $0.68\pi$ while lowering $V_\mathrm{tg}$, which well accounts for the phase shifts in Fig.~\ref{lowfield}(a). Particularly, the Berry phase in region T crosses $\pi$ at $V_\mathrm{tg}=-1.86\unit{V}$, which is consistent with the $\pi$-shift position in Fig.~\ref{lowfield}(a). Besides, the Berry phase in region B (\smash{$\Phi^\mathrm{\!\mathsmaller{ (B)}}_\mathrm{\!\mathsmaller{Berry}}$}) is only affected by $V_\mathrm{bg}$ and takes a constant value of $0.28\pi$ for $V_\mathrm{bg}=20\unit{V}$. 

The quasiparticle tunneling in gapped BLG is simultaneously tuned as the Berry phase changes in T. Given the variation of \smash{$\Phi^\mathrm{\!\mathsmaller{ (T)}}_\mathrm{\!\mathsmaller{Berry}}$}, we expect a transition from anti-Klein tunneling, corresponding to the Berry phase of $2\pi$, to Klein tunneling, at the Berry phase $\pi$, to reentrant anti-Klein tunneling upon further decreasing the Berry phase~\cite{varlet2015band}. To demonstrate the anticipated transitions, 
the transmission probability at normal incidence $T(\phi=0)$ (see Fig.~\ref{lowfield}(i)) is investigated by quantum transport simulations for two cases: phase-coherent (grey curve) and phase-incoherent (black curve). The phase-coherent transmission probability oscillates due to the resonance condition. Instead, the phase-incoherent transmission probability suppresses the resonance and is calculated by the relation $T(\phi=0)=1/(1/T_\mathrm{L}+1/T_\mathrm{R}-1)$~\cite{datta1997electronic}, where $T_\mathrm{L}$ and $T_\mathrm{R}$ represent the transmission probability through the left and right $pn$ interfaces of the potential barrier, respectively. The resulting phase-incoherent $T(\phi=0)$ agrees with our expectation except two differences.
(i) The Berry phase for anti-Klein tunneling, \smash{$\Phi^\mathrm{\!\mathsmaller{ (T)}}_\mathrm{\!\mathsmaller{Berry}}=2\pi$}, appears at the CNP, where zero charge carrier density in region T also gives rise to the inhibition of transmission as anti-Klein tunneling. (ii) The maximum $T(\phi=0)$ reaches 0.87 at $V_\mathrm{tg}=-1.24\unit{V}$, which is close to the unity transmission probability for perfect Klein tunneling~\cite{katsnelson2006chiral, katsnelson2012graphene,Shytov2008,Young2008}. The factor that impedes the maximum $T(\phi=0)$ to reach 1, is the Berry phase in region B, which is far from $\pi$.
But perfect Klein tunneling requires the Berry phase to be $\pi$ in both T and B regions. In addition, the reduction of $T(\phi=0)$ for $V_\mathrm{tg}<-1.24\unit{V}$ suggests that anti-Klein tunneling is partially restored. Therefore, the quasiparticle tunneling undergoes two processes: reaching Klein tunneling and recovering anti-Klein tunneling. 

\begin{figure}[b]
	\includegraphics[width=82mm]{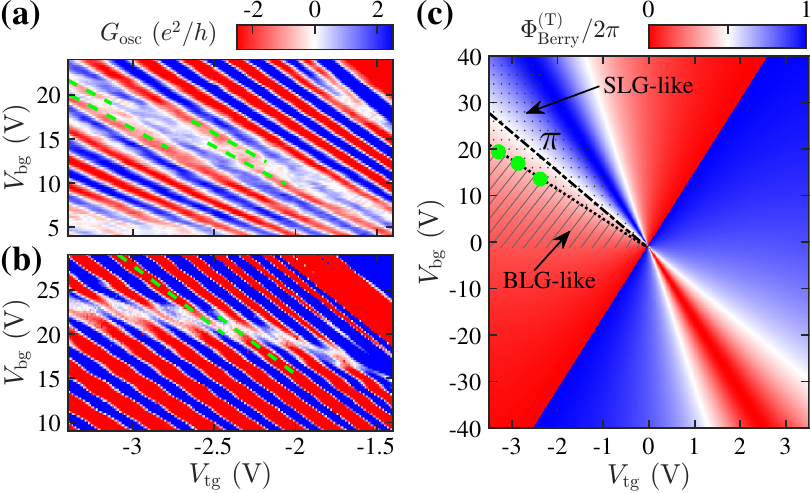}
	\caption{Fabry-P\'{e}rot interferences at zero magnetic field are shown both experimentally (a) and theoretically (b), zoomed in the white rectangle of Figs.~\ref{BL12D_FP_Gmap}(c) and (d), respectively. (c) Calculation of the Berry phase in region T as a function of $V_\mathrm{tg}$ and $V_\mathrm{bg}$. The dash-dotted line shows the position of \smash{$\Phi^\mathrm{\!\mathsmaller{(T)}}_\mathrm{\!\mathsmaller{Berry}}=\pi$}. The green dots mark the phase-shift positions appeared in (a), and define the dotted line.}
	\label{BL12D_FPshift}
\end{figure}

The transition from anti-Klein to Klein tunneling actually relies on the modulation of pseudospin orientation in gapped BLG. When the Fermi level is tuned close to the band edge, the pseudospin is rotated out of plane (see Fig.~\ref{AKtoK}(b)), leading to the broken chirality~\cite{min2008pseudospin,macdonald2012pseudospin,lundeberg2014harnessing,varlet2015band}. The momentum of charge carriers is, therefore, unlocked to the pseudospin, allowing Klein tunneling in gapped BLG (see Fig.~\ref{AKtoK}(f)). 
Even though the chirality sustains Klein tunneling in SLG~\cite{katsnelson2006chiral,katsnelson2012graphene}, the contrary happens in gapped BLG, \textit{i.e.}, Klein tunneling favors the impaired chirality. On the other hand, the chirality can be restored in gapped BLG~\cite{varlet2015band}, as long as the pseudospin recovers its in-plane orientation at sufficiently high Fermi energies; at the same time, the Berry phase of $2\pi$ (or equivalently $0$) as well as anti-Klein tunneling are regained. The recovery of anti-Klein tunneling is affected by two parameters, namely, the interlayer asymmetry and the Fermi energy. The chirality is broken because of the increasing interlayer asymmetry but recovered due to the rising Fermi energy.

\paragraph{SLG-like and BLG-like Berry phase.}

Figures \ref{BL12D_FPshift}(a) and (b) show the FP interference patterns magnified from the white rectangles in Figs.~\ref{BL12D_FP_Gmap}(c) and (d), respectively. 
Both experiments and simulation show nearly half-period shifts of the FP fringes, for example, highlighted by the green-dashed lines in Figs.~\ref{BL12D_FPshift}(a) and (b). The nearly half-period shifts indicate that a phase change of about $\pi$ is suddenly incorporated in the phase difference $\Delta\Phi$. This phase shifts can be attributed to two different ways, \textit{i.e.}, the SLG-like and the BLG-like, to acquire the Berry phase in gapped BLG. For better interpretation, we calculate the Berry phase in region T as a function of $V_\mathrm{tg}$ and $V_\mathrm{bg}$~\cite{varlet2014fabry} (see Fig.~\ref{BL12D_FPshift}(c)). The phase-shift positions in Fig.~\ref{BL12D_FPshift}(a) are labeled as three green dots on Fig.~\ref{BL12D_FPshift}(c), which arrange along the dotted line. The BLG-like \smash{$\Phi^\mathrm{\!\mathsmaller{ (T)}}_\mathrm{\!\mathsmaller{Berry}}$} can be acquired at zero magnetic field, hence, it appears as the predicted values in the striped-shade region of Fig.~\ref{BL12D_FPshift}(c). However, the SLG-like \smash{$\Phi^\mathrm{\!\mathsmaller{ (T)}}_\mathrm{\!\mathsmaller{Berry}}$} needs low magnetic fields to develop, and is unavailable at $B=0$. Actually, \smash{$\Phi^\mathrm{\!\mathsmaller{ (T)}}_\mathrm{\!\mathsmaller{Berry}}$} in the dotted-shade region remains zero instead of the calculated value. Accordingly, the phase shifts about $\pi$ show up around the intersection between the SLG-like and BLG-like regions, \textit{i.e.}, the dotted line, and directly prove the existence of two mechanisms, SLG-like and BLG-like, to obtain the Berry phase in gapped BLG.

\paragraph{Conclusion.}

We have examined the quasiparticle tunneling as well as the related Berry phase in BLG using a Fabry-P\'{e}rot interferometer based on a dual-gated geometry. As the crystal inversion symmetry is broken by applying a displacement field, a full control of the Berry phase within the range $0.68\pi$--$2\pi$ is achieved by manipulating the Fermi energy of charge carriers. Two distinct ways to acquire the Berry phase, SLG-like and BLG-like, coexist and can be switched between each other. Consequently, the corresponding quasiparticle tunneling undergoes a transition from anti-Klein to almost complete Klein tunneling with a maximum transmission probability of 0.87 at normal incidence. Therefore, in gapped BLG, tuning from BLG-like anti-Klein tunneling to SLG-like Klein tunneling is reachable by appropriate electrical gating.

\begin{acknowledgments}
We acknowledge A.\ Varlet for fruitful discussions. Financial supports from the Deutsche Forschungsgemeinschaft (DFG) within SFB 689, project Ri 681-13/1 and the program of the Forschungsgro{\ss}ger\"{a}te 121384/17-1, the Helmholtz association through the program STN, as well as the DFG Center for Functional Nanostructures (CFN) are gratefully acknowledged. 
\end{acknowledgments}

\end{document}


\begin{CJK*}{UTF8}{}

\title{Supplemental Material for \\ \textquotedblleft Tuning anti-Klein to Klein tunneling in bilayer graphene\textquotedblright}
\author{Renjun Du (\CJKfamily{gbsn}{杜人君})}
\email{Renjun.Du@outlook.com}
\affiliation{Institute of Nanotechnology, Karlsruhe Institute of Technology (KIT), D-76021 Karlsruhe, Germany}

\author{Ming-Hao Liu (\CJKfamily{bsmi}{劉明豪})}
\email{minghao.liu@phys.ncku.edu.tw}
\affiliation{Institut f\"{u}r Theoretische Physik, Universit\"{a}t Regensburg, D-93040 Regensburg, Germany}

\affiliation{Department of Physics, National Cheng Kung University, Tainan 70101, Taiwan}

\author{Jens Mohrmann}
\affiliation{Institute of Nanotechnology, Karlsruhe Institute of Technology (KIT), D-76021 Karlsruhe, Germany}

\author{Fan Wu (\CJKfamily{gbsn}{吴凡})}
\affiliation{Institute of Nanotechnology, Karlsruhe Institute of Technology (KIT), D-76021 Karlsruhe, Germany}
\affiliation{College of Optoelectronic Science and Engineering, National University of Defense Technology, Changsha 410073, China}

\author{Ralph Krupke}
\affiliation{Institute of Nanotechnology, Karlsruhe Institute of Technology (KIT), D-76021 Karlsruhe, Germany}
\affiliation{Institute of Material Science, Technische Universit\"{a}t Darmstadt, D-64287 Darmstadt, Germany}

\author{Hilbert v. L\"{o}hneysen}
\affiliation{Institute of Nanotechnology, Karlsruhe Institute of Technology (KIT), D-76021 Karlsruhe, Germany}
\affiliation{Institute for Solid State Physics and Physics Institute, Karlsruhe Institute of Technology (KIT), D-76021 Karlsruhe, Germany}

\author{Klaus Richter}
\affiliation{Institut f\"{u}r Theoretische Physik, Universit\"{a}t Regensburg, D-93040 Regensburg, Germany}

\author{Romain Danneau}
\email{romain.danneau@kit.edu}
\affiliation{Institute of Nanotechnology, Karlsruhe Institute of Technology (KIT), D-76021 Karlsruhe, Germany}

\date{\today}

\maketitle

\end{CJK*}

\section{Sample information}
\begin{figure}[htb]
	{\includegraphics[width=15cm]{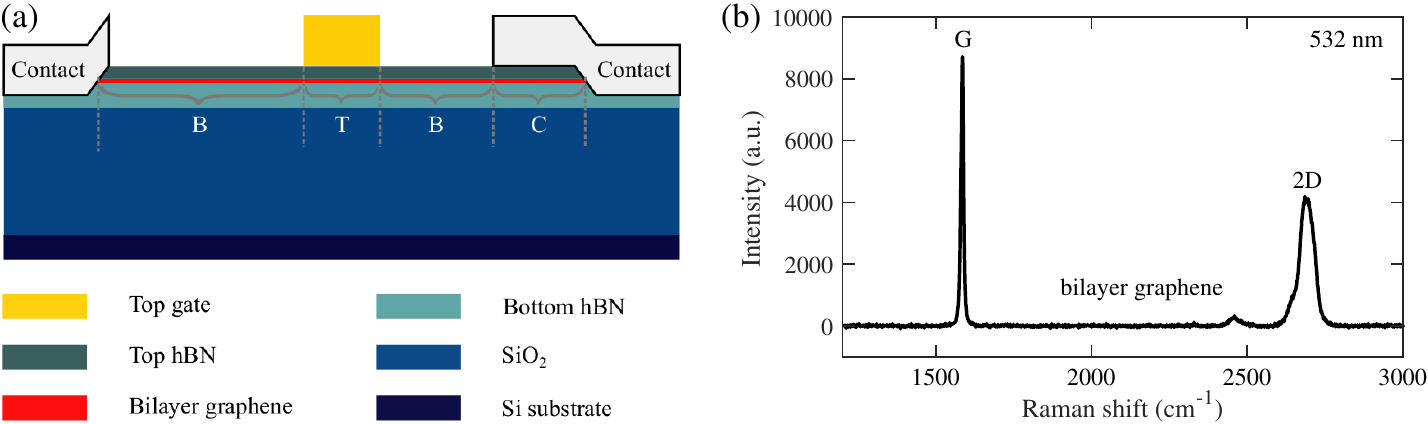}}
	{\caption{(a) Cross-section of the investigated devices. (b) Characterization of bilayer graphene: Raman spectrum. Laser wavelength 532 nm, power 2.5 mW, measurement duration 20 s.}
		\label{crosssketch}}
\end{figure}

In this section, we describe the configuration and geometry of the investigated devices.
Fig.~\ref{crosssketch}(a) shows the cross-section of the devices.
Bilayer graphene (BLG) or hexagonal boron nitride (hBN) flakes are first exfoliated on substrates with a scotch-tape technique. 
BLG is characterized by a RENISHAW inVia Raman spectrometer at a wave length of 532 nm (see Fig.~\ref{crosssketch}(b)).
The flatness and thickness of hBN flakes are measured by an atomic force microscope (AFM).
We then encapsulate BLG between top and bottom hBN flakes, which have the thickness of $14.5\unit{nm}$ and $35\unit{nm}$, respectively.
The hBN-BLG-hBN heterostructure is placed on Si substrate with a 317-nm thick SiO$_\mathrm{2}$ surface layer. 
Since the dielectric constant of SiO$_\mathrm{2}$ is known to be 3.9, we obtain the dielectric constant for hBN by fitting the slope of displacement field axis, i.e., $C_\mathrm{tg}/C_\mathrm{bg}\approx15.3$, yielding $\epsilon_\mathrm{r}^\text{hBN}\approx2.2$. 
We employ e-beam lithography to define a top gate (Ti/Au $5/75\unit{nm}$) about $\sim150\unit{nm}$ wide, and use the Si substrate as a global back gate.
BLG is then contacted from two side edges with metallic leads (Ti/Al $5/85\unit{nm}$). 
The contact resistance is estimated from the minimum resistance $R_\mathrm{min}$ measured in a two-terminal configuration. 
After subtracting the ballistic resistance $R_\mathrm{Q}=h/(4e^{2}\text{int}(2W/\lambda_\mathrm{F}))$ [$W$ is the width of the samples and $\lambda_\mathrm{F}$ is the Fermi wavelength of charge carriers], we obtain a low contact resistivity $(R_\mathrm{min}-R_\mathrm{Q})W/2\approx60 \unit{\si{\ohm\micro\meter}}$~\cite{Wang2013a,calado2015ballistic,shalom2016quantum} due to the highly transparent metal-graphene interfaces.
All measurements are performed at $4.2\unit{K}$.

Two devices, investigated in this work, have the same $W/L$ ratio of 5 but slightly different channel lengths, i.e., $0.8 \unit{\si{\micro\meter}}$ for device PNJ-A and $1 \unit{\si{\micro\meter}}$ for device PNJ-B.
The channel consists of four parts,  marked as T (top- and back-gated region), B (only back-gated regions), and C (contact-overlapping region). 
The geometric parameters of each section are listed in Table~\ref{parameter}. The two devices were fabricated with same BLG on one substrate, and during the same run. However, their measurements were performed in different cool-downs. As a result of thermal cycling, the intrinsic doping of BLG changes.

\begin{table}[htb]
	\scriptsize
	\centering
	{\caption{The geometric informations of the devices PNJ-A and PNJ-B.}\label{parameter}}
		\begin{tabular}{m{2cm}m{1.5cm}m{1.5cm}m{1.5cm}m{1.5cm}m{1.5cm}}
				\hline\hline
		\multirow{2}{*}{Sample} &  {Width (\si{\micro\meter})} & \multicolumn{4}{c}{Length (nm)} \\
											 &            &    {B} & {T} & {B} & {C}  \\\hline
		PNJ-A &  4 & 310 & 150 & 157 & 181  \\
		PNJ-B &  5 &  427 & 157 & 235 & 176 \\
		\hline
		\multirow{2}{*}{Thickness (nm) } &
		{ Bottom hBN}	& {Top hBN} & {SiO$_{2}$} & {Contacts} &	{Top gate} \\\cline{2-6}
		&  35 &  14.5  &  317  & Ti/Al 5/85 &  Ti/Au 5/75  \\ \hline\hline
	\end{tabular}
\end{table}

\section{Electrostatic model}
\begin{figure}[b]
	\includegraphics[width=0.47\textwidth,page=1]{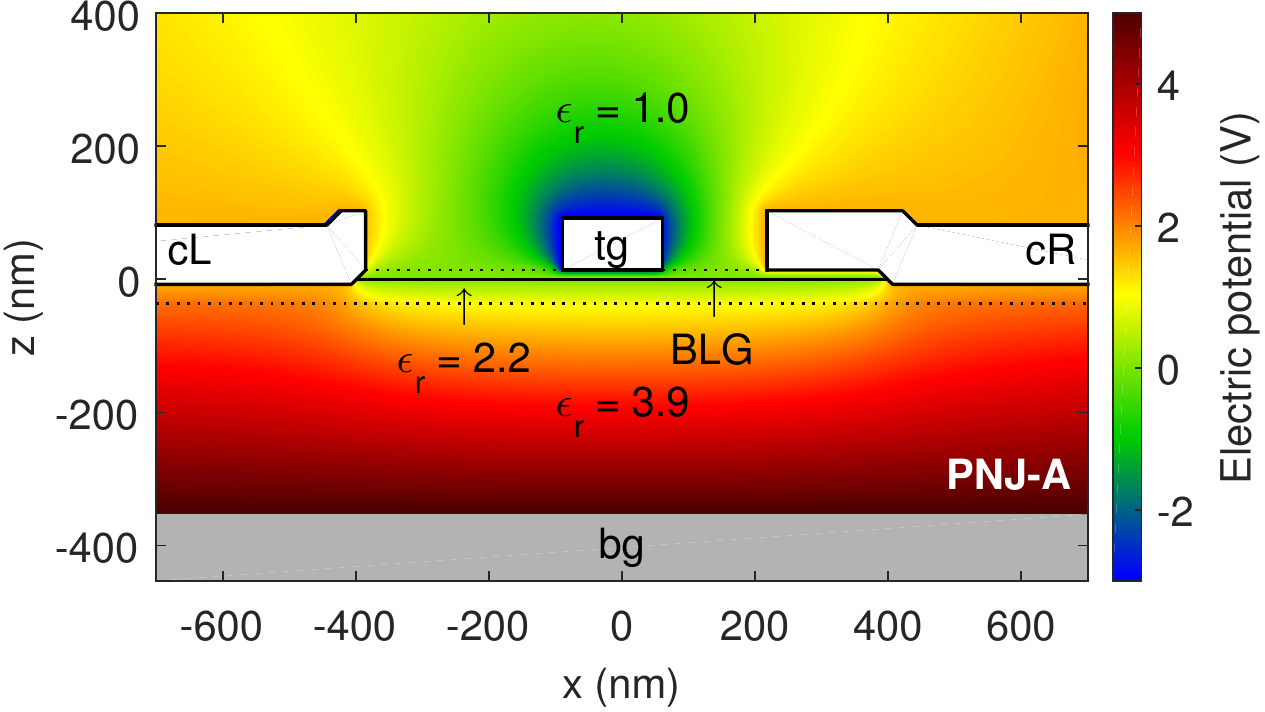}\hfill\includegraphics[width=0.47\textwidth,page=2]{FigS2.pdf}
	\caption{Geometry of the 2D electrostatic models for the devices PNJ-A and PNJ-B. Exemplary electric potential distributions are obtained by electrostatic simulations, considering gate voltages $(V_\mathrm{tg},V_\mathrm{bg}) = (-3,5)\unit{V}$ in both devices and effective contact doping potential $V_\mathrm{cL}=V_\mathrm{cR}=1.6\unit{V}$ for PNJ-A and $1.1\unit{V}$ for PNJ-B.}
	\label{fig Esim}
\end{figure}

Following the experimentally measured device geometry, we construct 2D electrostatic models for PNJ-A and PNJ-B as shown in Fig.~\ref{fig Esim}, where exemplary electrostatic potential distributions are obtained by finite-element-based electrostatic simulation using \textsc{FEniCS} \cite{fenics} combined with the mesh generator \textsc{Gmsh} \cite{gmsh}. The BLG, together with four metallic electrodes in each model --- left contact (cL), right contact (cR), top-gate (tg) and back-gate (bg) --- form a linear system of coupled conductors. The carrier density of the BLG sample is given by
\begin{equation}
	n(x) = \sum_i \frac{C_i(x)}{e}V_i\ +n_0(x),
	\label{eq n}
\end{equation}
where the electrode label index runs over $i=\mathrm{cL,cR,tg,bg}$, and the self-partial capacitance $C_i$ can be obtained from the electrostatic simulation by treating BLG as the reference conductor \cite{Cheng1989,Liu2013}. Note that in Eq.~\eqref{eq n}, gate potentials $V_\mathrm{tg}$ and $V_\mathrm{bg}$ directly correspond to the top- and back-gate voltages in experiments, respectively, while effective potentials $V_\mathrm{cL}$ and $V_\mathrm{cR}$ due to contact doping \cite{Liu2013a}, together with the intrinsic doping $n_0(x)$, are to be determined by analyzing the conductance measurement from the experiment, as illustrated in the following.

\begin{figure}[b]
	\subfigure[]{
		\includegraphics[width=0.48\textwidth]{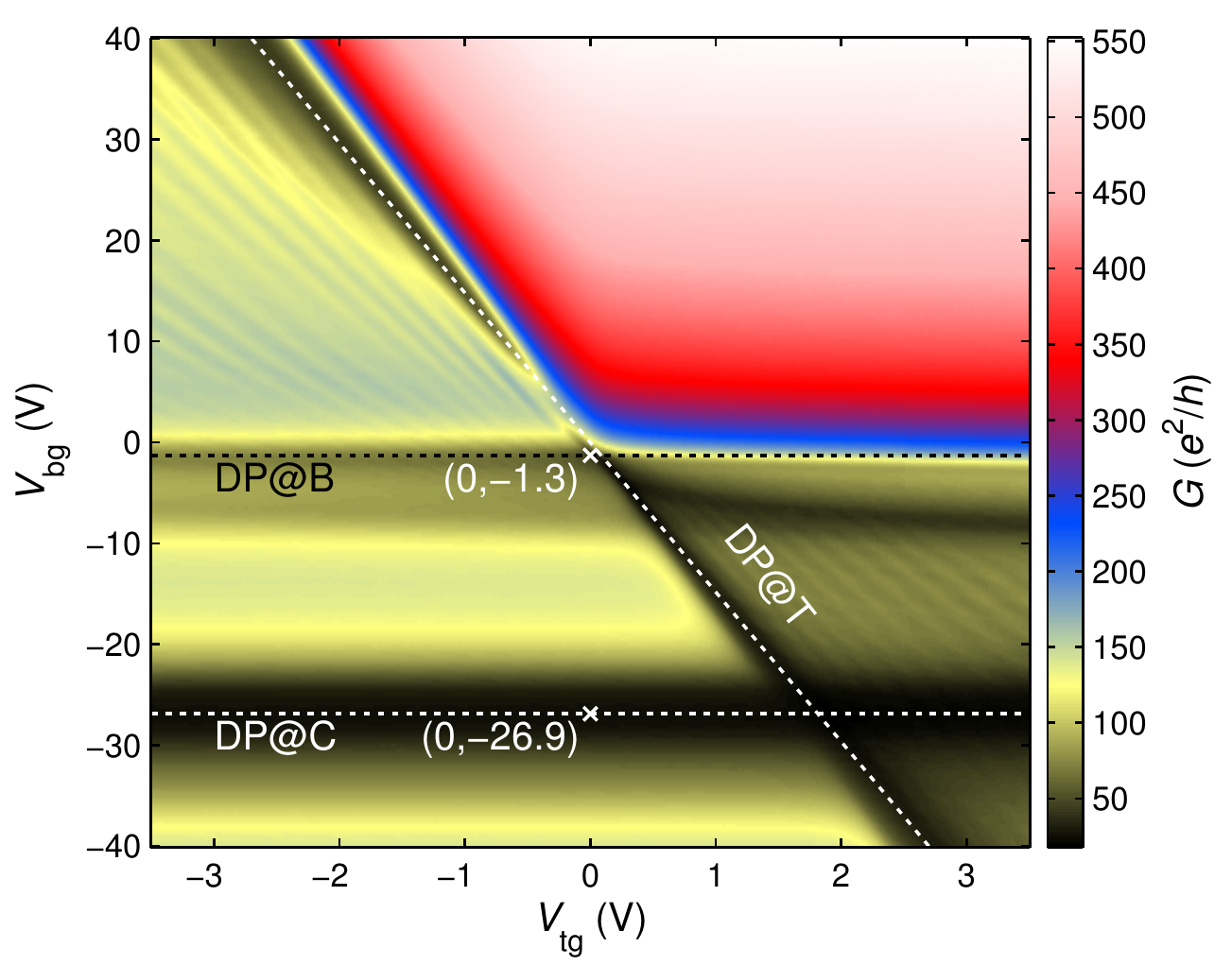}}\hfill \subfigure[]{
		\includegraphics[width=0.48\textwidth]{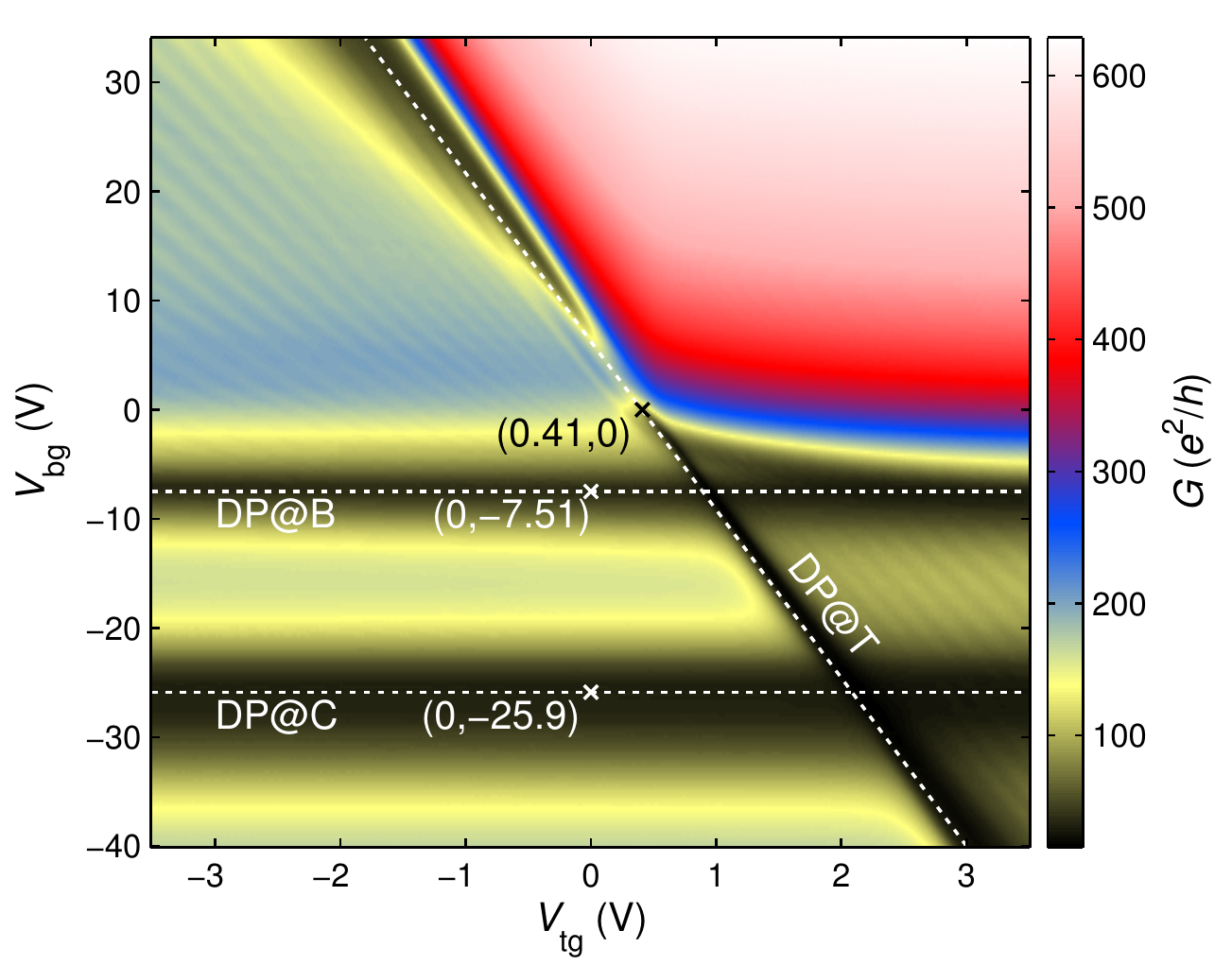}}
	\caption{Analysis of the Dirac point (DP) in different regions of the BLG sample for (a) PNJ-A and (b) PNJ-B from the conductance measurement. Three DP axes indicated by the dashed lines are identified in each panel: DP@B standing for DP in the B region, etc.}
	\label{figS GVVexp}
\end{figure}

\subsection{Contact and intrinsic doping}\label{sec doping}

To deduce the effective contact doping potential and the intrinsic doping, i.e., $V_\mathrm{cL},V_\mathrm{cR}$ and $n_0(x)$ in Eq.~\eqref{eq n}, we infer the conductance measured as a function of top- and back-gate voltages at zero magnetic field (see Fig.~\ref{figS GVVexp}), and analyze individually for the two devices in the following.

\paragraph{PNJ-A}

\begin{figure}[t]
	\subfigure[\ PNJ-A]{\includegraphics[height=4cm,page=1]{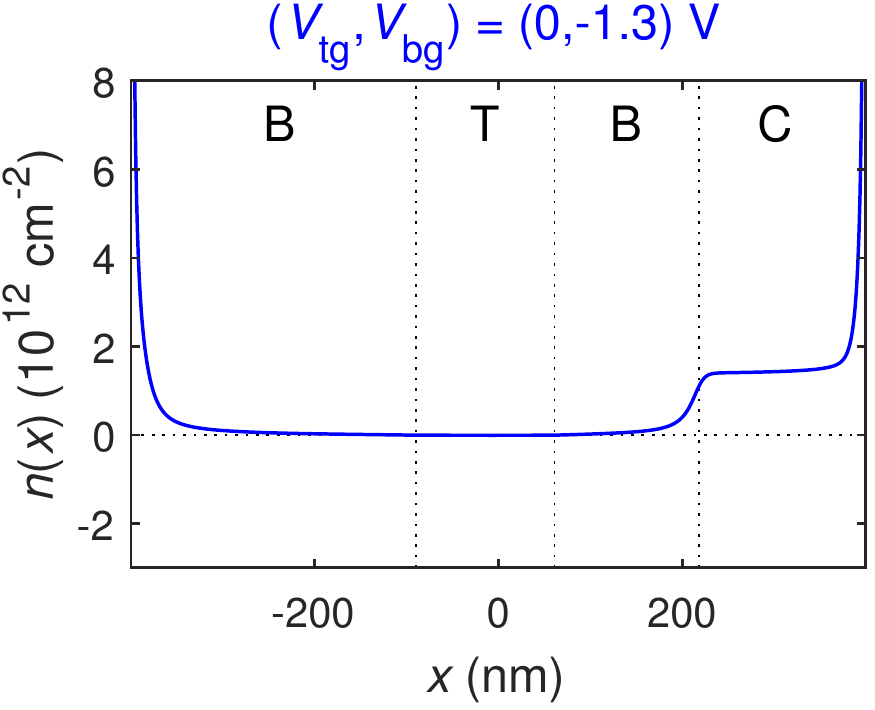}\quad \includegraphics[height=4cm,page=2]{FigS4_DPanalysis_A.pdf}}
	
	\subfigure[\ PNJ-B]{\includegraphics[height=4cm,page=1]{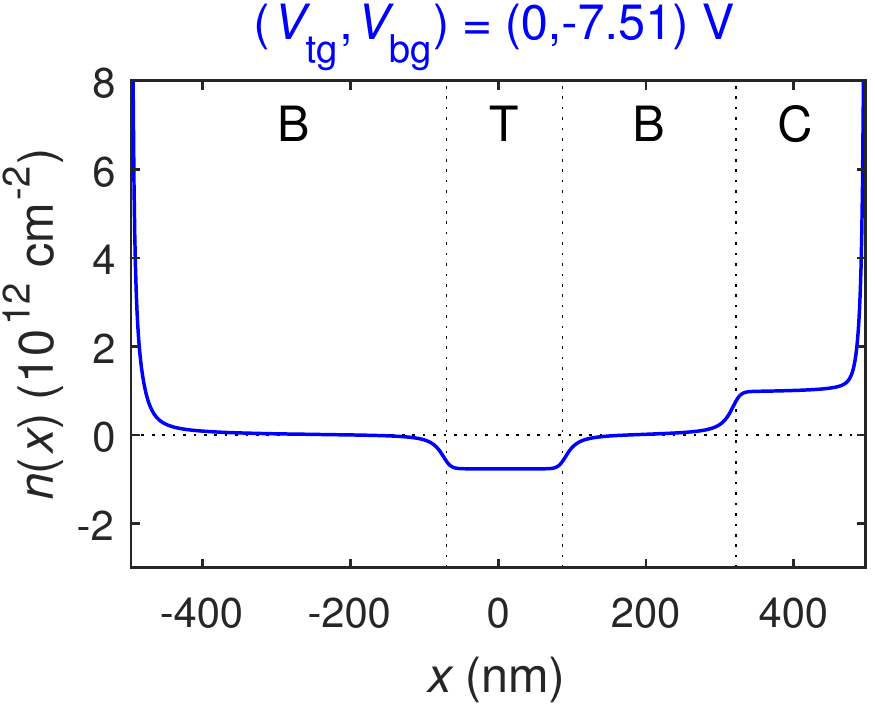}\quad \includegraphics[height=4cm,page=2]{FigS4_DPanalysis_B.pdf}\quad \includegraphics[height=4cm,page=3]{FigS4_DPanalysis_B.pdf}}
	\caption{Carrier density profiles as a consistency check confirming the contact and intrinsic doping models for PNJ-A and PNJ-B deduced in \autoref{sec doping}.}
	\label{figS nx DP check}
\end{figure}

From the Dirac point in region B (DP@B) at $V_\mathsf{bg}=-1.3\unit{V}$ shown in Fig.~\ref{figS GVVexp}(a), a global doping concentration of
\begin{equation}
	n_0 = \frac{C_\mathsf{bg}}{e}\times 1.3\unit{V} = 6\times 10^{10}\unit{cm}^{-2}
	\label{eq n0 for PNJ-A}
\end{equation}
can be deduced. For DP@C, we numerically found that the charge neutrality in the C region at $V_\mathrm{bg}=-26.9\unit{V}$ can be reached by setting
\begin{equation}
	V_\mathsf{cR}=V_\mathsf{cL}=1.6\unit{V}.
	\label{eq Vc for PNJ-A}
\end{equation}
Parameters \eqref{eq n0 for PNJ-A} and \eqref{eq Vc for PNJ-A} required in Eq.~\eqref{eq n} complete the electrostatic model for PNJ-A, and are confirmed numerically in Fig.~\ref{figS nx DP check}(a). Note that a horizontal low-conductance axis below DP@B can be vaguely seen, but is neglected here for simplicity. Together with the fact that the DP@B and DP@T lines in Fig.~\ref{figS GVVexp}(a) cross each other nearly at $(0,0)$, a constant $n_0$ seems sufficient to provide a satisfactory electrostatic model. In the following, we will show that a position-dependent $n_0(x)$ model is required for PNJ-B.

\paragraph{PNJ-B}

From the point $(0,-7.51)$ on the DP@B axis of Fig.~\ref{figS GVVexp}(b), we can deduce a global doping concentration to be
\begin{equation}
	n_0 = \frac{C_\mathsf{bg}}{e}\times 7.51\unit{V} = 4.26\times 10^{11}\unit{cm}^{-2}\ ,
	\label{eq n0 for PNJ-B}
\end{equation}
which allows us to reproduce the DP@B axis. Similar to PNJ-A, we found that setting
\begin{equation}
	V_\mathsf{cR}=V_\mathsf{cL}=1.1\unit{V}
	\label{eq Vc for PNJ-B}
\end{equation}
allows us to reach charge neutrality in the C-region and hence reproduce the DP@C axis. For the DP@T axis, however, the fact that the point $(0.41,0)$ on the DP@T axis is quite distant from $(0,-7.51)$ on the DP@B axis, suggests a nonuniform doping centered in the T region. To minimize the introduction of additional parameters, let us assume the nonuniform intrinsic doping centered at the T region, called $n_0^T(x)$, to have the same profile as the top-gate capacitance $C_\mathsf{tg}(x)$. Let:
\begin{equation}
	n_0^T(x)\equiv \frac{C_\mathsf{tg}(x)}{e}V_\mathsf{tg}^0\ ,
	\label{eq n0T}
\end{equation}
where $V_\mathrm{tg}^0$ is the only additional parameter introduced, allowing us to cast $n_0^T(x)$ into a shift of the top-gate voltage, i.e., Eq.~\eqref{eq n} can be now written as
\begin{equation}
	n(x) = \frac{C_\mathsf{tg}(x)}{e}\left(V_\mathsf{tg}+V_\mathsf{tg}^0\right)+\frac{C_\mathsf{bg}(x)}{e}V_\mathsf{bg}+\frac{C_\mathsf{cL}(x)}{e}V_\mathsf{cL}+\frac{C_\mathsf{cR}(x)}{e}V_\mathsf{cR}+n_0\ .
	\label{eq n(x) PNJ-B}
\end{equation}
By considering the carrier density in the center of cavity T, say $x_T$, at the voltage point $(0.41,0)$ on the DP@T axis of Fig.~\ref{figS GVVexp}(b), we obtain:
\begin{equation}
	n(x_T)\approx \frac{C_\mathsf{tg}(x_T)}{e}\left(0.41\unit{V}+V_\mathsf{tg}^0\right)+n_0=0 \implies V_\mathsf{tg}^0 = -0.89\unit{V}\ .
	\label{eq Vtg0}
\end{equation}
Model function \eqref{eq n0T} with the parameter $V_\mathrm{tg}^\mathrm{0}$ \eqref{eq Vtg0}, together with the uniform part of the intrinsic doping \eqref{eq n0 for PNJ-B} and the contact doping potential \eqref{eq Vc for PNJ-B}, complete the electrostatic model using \eqref{eq n}, which is explicitly written as \eqref{eq n(x) PNJ-B} for PNJ-B. As a consistency check, the above doping model is numerically confirmed in Fig.~\ref{figS nx DP check}(b).

\subsection{Examples of carrier density profiles}

\begin{figure}
	\subfigure[\ PNJ-A]{
		\includegraphics[height=4cm,page=2]{FigS5_nx_A.pdf}}\quad\subfigure[\ PNJ-B]{
		\includegraphics[height=4cm,page=2]{FigS5_nx_B.pdf}}\hfill
	\subfigure[\ PNJ-B]{
		\includegraphics[height=4cm,page=4]{FigS5_nx_B.pdf}}
	
	\subfigure[\ PNJ-A]{
		\includegraphics[height=4cm,page=1]{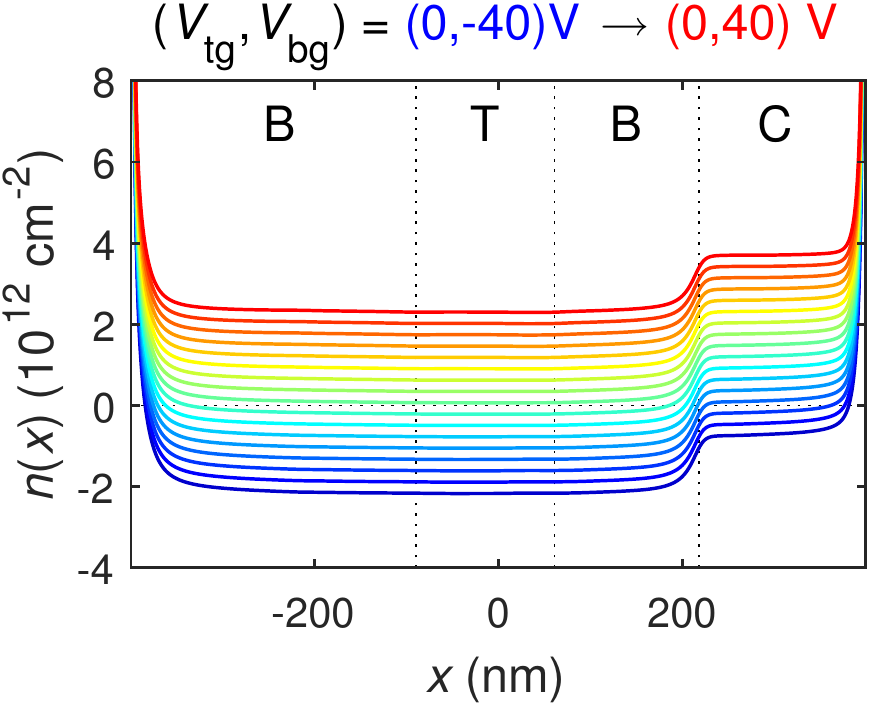}}\quad\subfigure[\ PNJ-B]{
		\includegraphics[height=4cm,page=1]{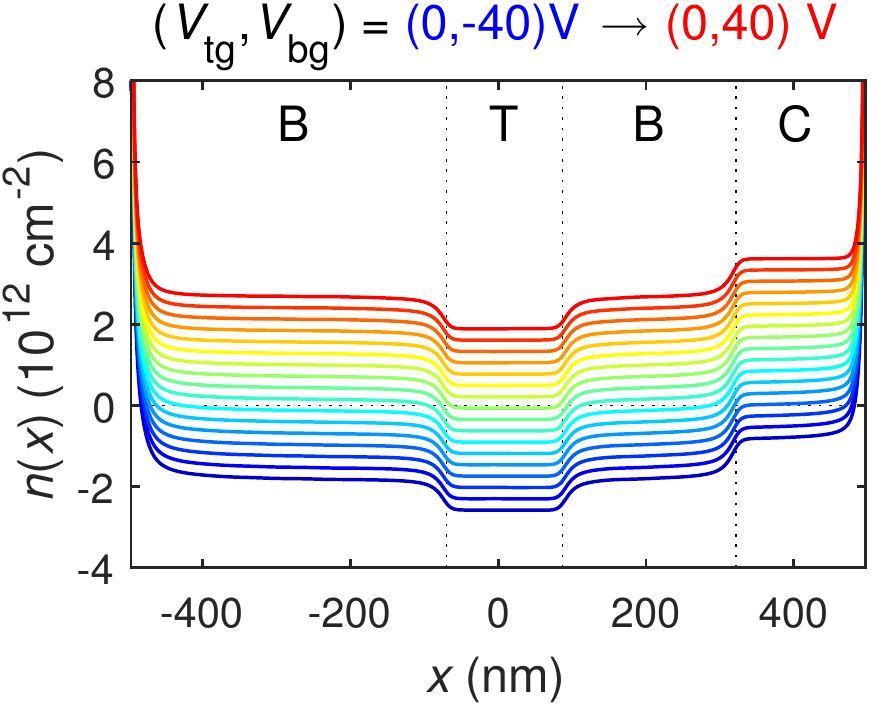}}\hfill
	\subfigure[\ PNJ-B]{
		\includegraphics[height=4cm,page=3]{FigS5_nx_B.pdf}}
	
	\caption{Examples of carrier density profiles from the electrostatic simulation combined with the deduced contact and intrinsic doping described in \autoref{sec doping}, showing (a)--(c) top-gate and (d)--(f) back-gate sweeps. Device labels are indicated in each subfigure label. Panels (a,b,d,e) are basic characterizations for both devices. Further examples for PNJ-B are shown in (c) for the top-gate sweep considered in Fig.~3 of the main text and (d) for the back-gate sweep considered in Fig.~\ref{S_BL12C_FP_Gmap}(g).}
	\label{fig n(x)}
\end{figure}

Based on the models introduced above, we show examples of simulated carrier-density profiles in Fig.~\ref{fig n(x)}, considering top-gate sweeps with back gate grounded in panel (a)/(b) and back-gate sweeps with top gate grounded in panel (d)/(e) for device PNJ-A/B.

Additional examples for PNJ-B are shown in Fig.~\ref{fig n(x)}(c) and (f). The former corresponds to the top-gate sweep with back gate fixed at $V_\mathrm{bg}=20\unit{V}$ considered in Fig.~3 of the main text, while the latter the back-gate sweep with top gate fixed at $V_\mathrm{tg}=-1.5\unit{V}$ considered in Fig.~\ref{S_BL12C_FP_Gmap}(g). 

\subsection{Local band offset (on-site energy) profiles}

To implement the carrier density profiles $n(x)$ from the electrostatic simulation in transport calculations based on the tight-binding model, we need to translate $n(x)$ into the local band-offset profile $V(x)$ (also known as the on-site energy profile), i.e., the diagonal elements in the site-resolved tight-binding Hamiltonian:
\[
H = H_0+\sum_i V(x_i) c_i^\dag c_i\ .
\]
In the above expression, the first term $H_0$ is the pristine part of the BLG Hamiltonian composed only hopping elements, and the second term is the on-site energy with the site index $i$ running over all sites within the considered scattering region and $V(x_i)$ being the energy offset applied on site $i$.

As already mentioned in the main text, the simulation scheme of the present work is basically the same as that in \cite{varlet2014fabry}, where a simplified model is considered such that in each region of the simulated BLG device, the carrier density is position-independent, and so are the corresponding asymmetry parameters and on-site energies. In the Supplemental Material (SM) of \cite{varlet2014fabry}, it was shown how the on-site energy $V$ can be obtained from the gate-controlled carrier density $n$ taking into account the asymmetry parameter $U$ \cite{mccann2013electronic}: 
\[
\tikzpicture
\node (n) at (0,0) {$n$}; 
\node (U) at (1.5,0) {$U$};
\node (V) at (3,0) {$V$}; 
\draw [->] (n) -- (U);
\draw [->] (n) -- +(0,0.7) -| (V);
\draw [->] (U) -- (V);
\endtikzpicture
\]
In the present work, we adopt the same method but keep the position dependence of the simulated carrier density $n(x)$, and hence of the corresponding asymmetry parameter $U(x)$ and the resulting on-site energy profile $V(x)$:
\[
\tikzpicture
\node (n) at (0,0) {$n(x)$}; 
\node (U) at (1.5,0) {$U(x)$};
\node (V) at (3,0) {$V(x)$}; 
\draw [->] (n) -- (U);
\draw [->] (n) -- +(0,0.7) -| (V);
\draw [->] (U) -- (V);
\endtikzpicture
\]

\begin{figure}
	\includegraphics[width=14cm]{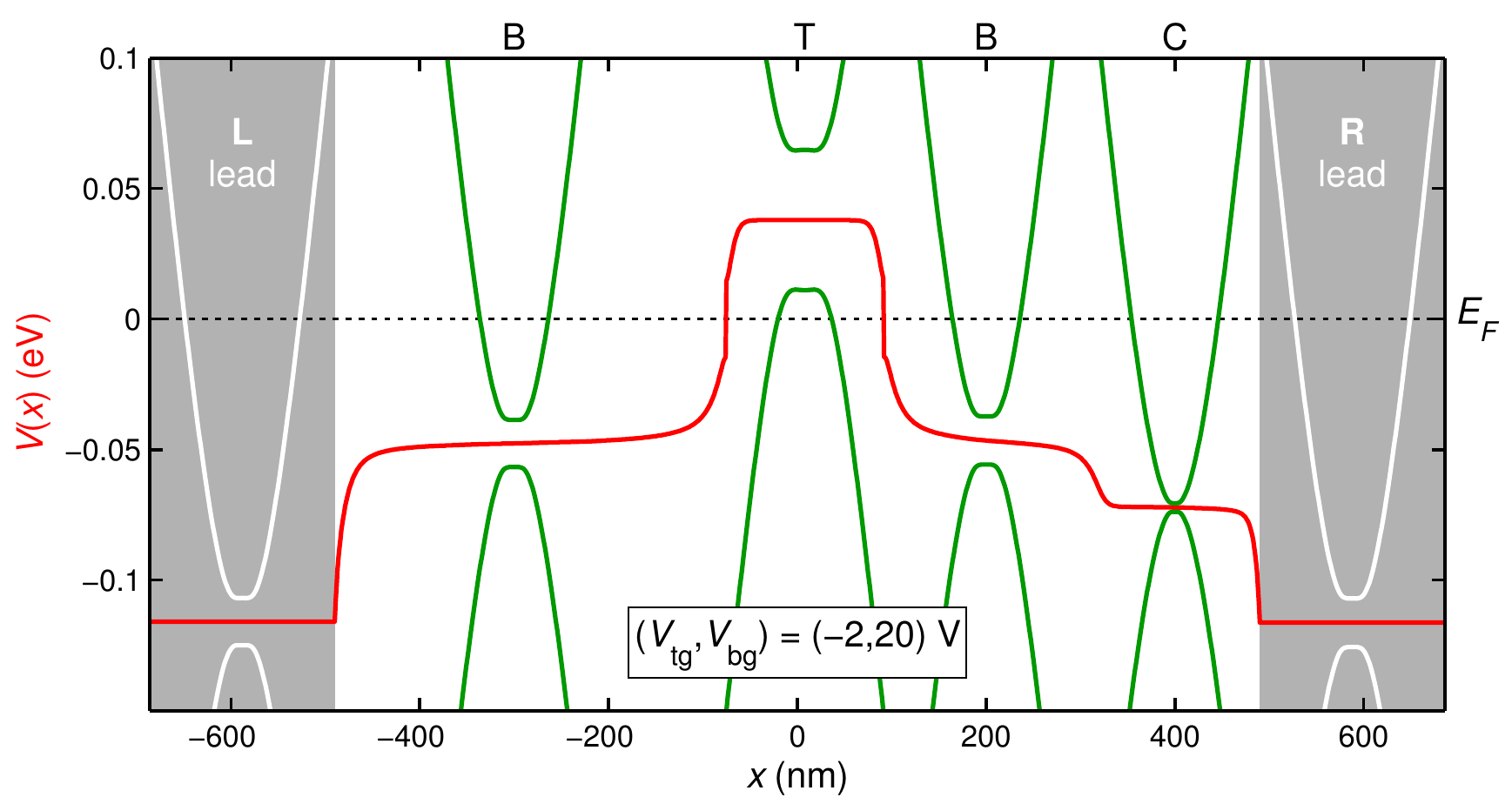}
	\caption{An example of the on-site energy profile $V(x)$ inferred from the simulated carrier density profile $n(x)$, considering the device PNJ-B at gate voltages indicated above. Local band structures in the B, T, and C regions are sketched in green with the position-dependent band gap [$\approx$ the asymmetry parameter $U(x)$] implemented.}
	\label{figS Vx with BS}
\end{figure}

In Fig.~\ref{figS Vx with BS}, we show an example of the resulting on-site energy profile, together with local band structures with the position-dependent $U(x)$ implemented and properly offset by $V(x)$, considering PNJ-B with gate voltages $(V_\mathrm{tg},V_\mathrm{bg}) = (-2,20)\unit{V}$, which is around the center point of Fig.~3(b) of the main text. From the local band structures shown in Fig.~\ref{figS Vx with BS}, it can be seen that the Fermi level is close to the band edge in the T region but far away in the B region. Thus the Berry phase in T is expected to be close to $\pi$ but close to $0$ (or equivalently $2\pi$) in B, like in gapless BLG.

\section{Conductance calculation}
Quantum transport simulations are performed using a real-space Green's function method~\cite{liu2012efficient} based on the tight-binding model for Bernal-stacked BLG~\cite{PhysRev1957McClure,PhysRev1958Slonczewski}. The main task is to compute the single-mode conductance $g$, which is obtained by integration of the transmission function over all incident angles~\cite{liu2012efficient} and normalized to $e^2/h$.The resulting $g$ ranges from 0 to 2 since the spin degeneracy is not taken into consideration but the valley degeneracy is embedded in the tight-binding Hamiltonian.
When considering the conduction modes $M$ and the contact resistance $R_{\rm{c}}$, the full conductance $G$ is derived from the normalized single-mode $g$ by the following relation~\cite{Rickhaus2013}
\begin{equation}
G=\frac{2e^{2}}{h}\left(\frac{1}{Mg}+\frac{R_{\rm{c}}}{h/2e^{2}} \right)^{-1}.
\end{equation}  
We obtain $R_{\rm{c}}=1.1622\times10^{-3}\frac{h}{2e^{2}}$ for device PNJ-A and $R_{\rm{c}}=1.5099\times10^{-3}\frac{h}{2e^{2}}$ for device PNJ-B. 
Also, $M$ can be extracted by $M=W\sqrt{\overline{n}/\pi}$, where $\overline{n}=\overline{|n(x)|}$ is the absolute charge carrier density averaged over the entire scattering region and $W$ is the width of the channel. 
Since $R_{\rm{c}}$ and $M$ are determined  unequivocally, $G$ is calculated properly.

\section{Fabry-P\'{e}rot (FP) interferences}
\begin{figure}[b]
	{\includegraphics[width=15.6cm]{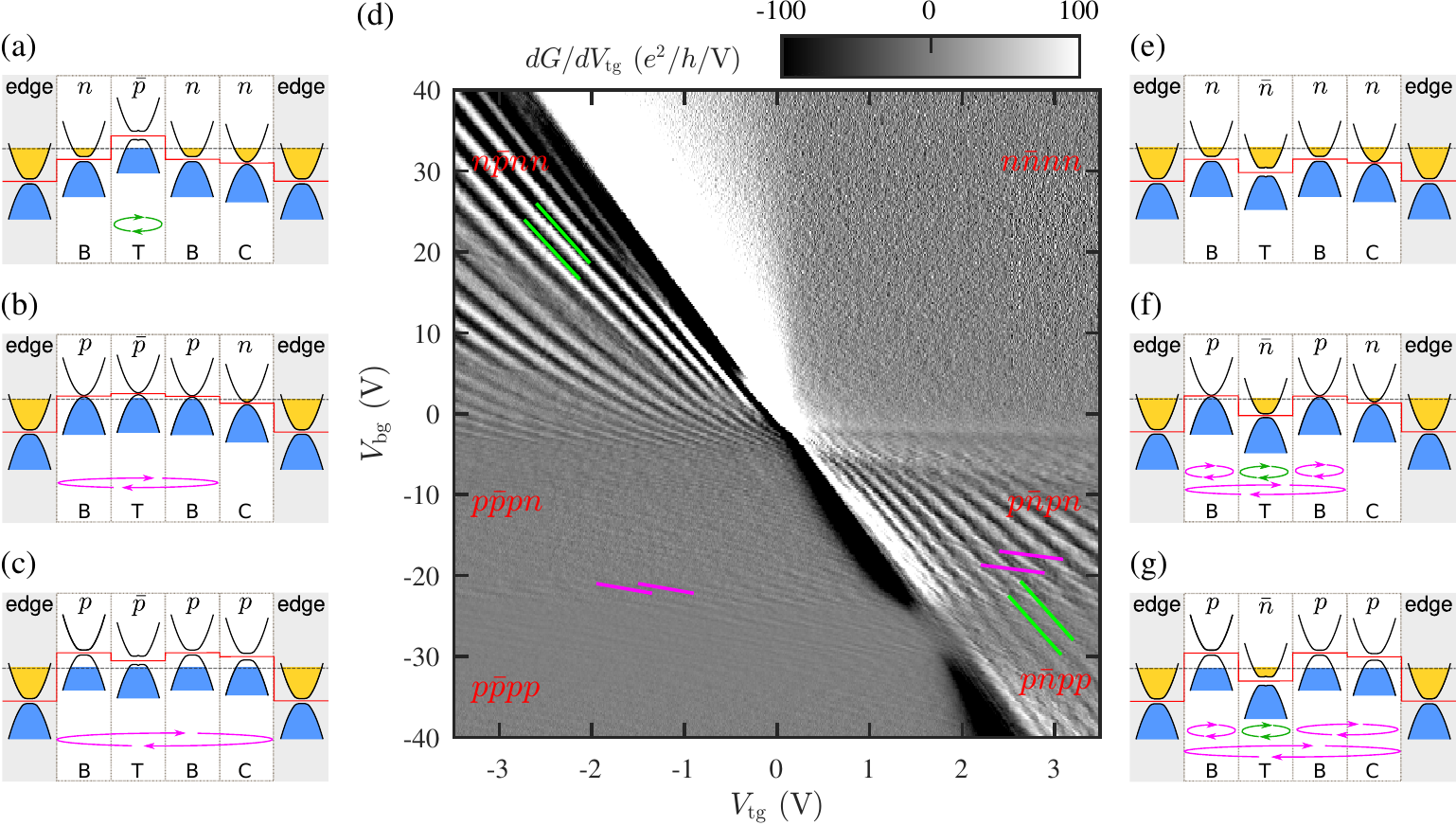}}
	{\caption{ Sketches of the potential profiles and cavities for the $n\bar{p}nn$ (a), $p\bar{p}pn$ (b), $p\bar{p}pp$ (c), $n\bar{n}nn$ (e), $p\bar{n}pn$ (f),  $p\bar{n}pp$ (g) regimes. The red solid line shows the potential profile. The gray dashed line indicates the Fermi level. The green arrows label the FP cavity affected by both top and back gates. The magenta arrows mark the back-gated cavities. (d) Transconductance $dG/dV_\mathrm{tg}$ as a function of both $V_\mathrm{tg}$ and $V_\mathrm{bg}$ for device PNJ-A. The green lines indicate the orientation of the FP fringes tuned by both gates. The magenta lines follow the direction of the fringes tuned by the back gate.}
		\label{Bl12D_FPinterference}}
\end{figure}

FP interferences occur when a cavity is formed between two parallel semitransparent $pn$ interfaces of a potential barrier~\cite{Young2008,Rickhaus2013,varlet2014fabry}.
We create the potential barrier across the device by tuning $V_\mathrm{tg}$ and $V_\mathrm{bg}$.
For better visualizing FP interferences, we show the transconductance $dG/dV_\mathrm{tg}$ as a function of $V_\mathrm{tg}$ and $V_\mathrm{bg}$ for device PNJ-A (see Fig.~\ref{Bl12D_FPinterference}(d)). 
In the following, we analyze the FP fringes and their formation for each region.

\subsection{FP interferences in the unipolar regime}

\textbf{Resonance in the $n\bar{n}nn$ regime.} 
As presented in Fig.~\ref{Bl12D_FPinterference}(d), FP interferences are absent only in the $n\bar{n}nn$ regime.
This absence can be easily understood with the help of the sketch in Fig.~\ref{Bl12D_FPinterference}(e), which portrays the potential profile together with the band diagrams.
Except the four regions (B, T, B and C) of the channel, we also consider the two edges of the device, since these edges are strongly $n$-doped by the Ti/Al leads, and may lead to two extra  $pn$ interfaces at the edges.
In the case of the $n\bar{n}nn$ regime, the charge carriers are $n$-type across the device, therefore, a cavity cannot be generated.

\textbf{Resonance in the $p\bar{p}pp$ and $p\bar{p}pn$ regimes.} Weak conductance resonance is observed in the $p\bar{p}pp$ and $p\bar{p}pn$ regimes (see Fig.~\ref{Bl12D_FPinterference}(d)). 
The fringes stretch out as marked with the magenta lines. 
Even though these fringes slightly lean against the horizontal line (CNP) owing to the effect of $V_\mathrm{tg}$, $dG/dV_\mathrm{tg}$ primarily oscillates as $V_\mathrm{bg}$ varies. 
We ascribe these fringes to FP interferences in cavities modulated individually by $V_\mathrm{bg}$. 
These cavities develop due to the formation of extra p-n interfaces at the edges of the device, as the magenta arrows shown in Figs.~\ref{Bl12D_FPinterference}(b)--(c). 
For instance, the entire device becomes a large cavity in the $p\bar{p}pp$ regime (see Fig.~\ref{Bl12D_FPinterference}(c)).
Since these cavities are large, both the amplitudes and periods of the resonance are reduced. 

\subsection{FP interferences in the bipolar regime}

\textbf{Resonance in the $n\bar{p}nn$ regime.} 
In the bipolar regime, the FP fringes follow the direction of the diagonal line, marked by the green lines in Fig.~\ref{Bl12D_FPinterference}(d). 
Thus, the resonance arises from a cavity created in T region, labeled by the green arrows in Fig.~\ref{Bl12D_FPinterference}(a). 
Due to the small cavity, about $150\unit{nm}$ wide, the resonance displays large periods and amplitudes.

The cavity length is extracted utilizing discrete Fourier transformation to analyze the measured conductance oscillations~\cite{varlet2014fabry}.
By transforming the oscillations from wave vector $k$ space ($k=\sqrt{\pi |n_{\rm{T}}|}$, $n_{\rm{T}}$ is charge carrier density in region T)  to frequency $\omega$ space, we determine the leading frequency of the conductance oscillations. 
For one period of oscillation, the phase difference satisfies $\Delta \Phi=\Delta k\cdot2L_{\rm{cavity}}=2\pi$, where $\Delta k$ denotes the change of wave vector in one period.
As the period is determined by discrete Fourier transformation, the cavity length is obtained.

\textbf{Resonance in the $p\bar{n}pn$ and $p\bar{n}pp$ regimes.} 
The FP patterns in the $p\bar{n}pn$ and $p\bar{n}pp$ regimes consist of two sets of fringes. 
The major one, dispersing along the green lines in Fig.~\ref{Bl12D_FPinterference}(d), originates from the same cavity T as in the $n\bar{p}nn$ regime (see the green arrows in Figs.~\ref{Bl12D_FPinterference}(f)--(g)). 
The fine resonance marked by the magenta lines, comes from the interferences in three cavities modulated only by the back-gate voltages, as shown by the magenta arrows in Figs.~\ref{Bl12D_FPinterference}(f)--(g). 
In the case of $p\bar{n}pn$ regime, the sizes of those three cavities are consistent with the lengths of the left B ($310\unit{nm}$), the right B ($157\unit{nm}$) and the entire B-T-B ($617\unit{nm}$) regions, as listed in Table \ref{parameter}.
In the case of $p\bar{n}pp$ regime, the three cavities are B (left), B-C, B-T-B-C, and the cavity lengths can be obtained from Table \ref{parameter}.
Although the fringe orientations are the same for those three cavities, the periods and amplitudes change due to the different cavity lengths. 

\section{Simulations of the Berry phase and quasiparticle tunneling for device PNJ-B}

\begin{figure}[t]
	{\includegraphics[width=13.5cm]{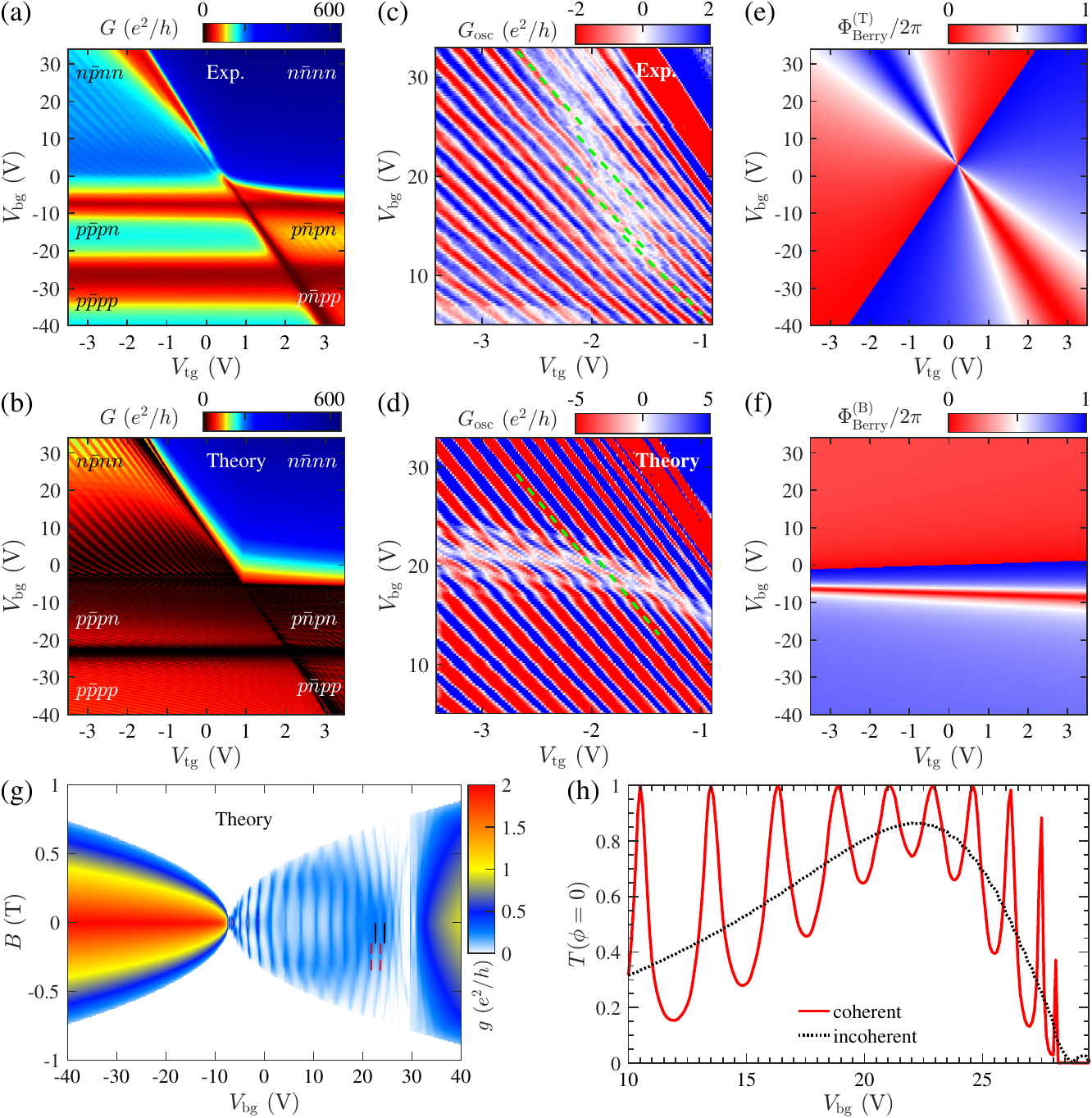}}
	{\caption{Conductance measurements (a) and the corresponding simulation result (b) for device PNJ-B at $B=0$. The oscillatory part of the conductance for the experimental (c) and simulated (d) results zoomed in the $n\bar{p}nn$ regime of  (a) and (b), respectively. The green dashed lines point out the positions of the phase shifts. The Berry phases in regions T (e) and B (f) evolve with $V_\mathrm{tg}$ and $V_\mathrm{bg}$ for device PNJ-B. (g) A numerical example of the single-mode conductance $g$ as a function of  $V_\mathrm{bg}$ and $B$ at $V_\mathrm{tg}=-1.5$ V. The black solid and red dashed lines show the initial and shifted positions of the FP fringes, respectively. (h) The associated transmission probability at normal incidence $T(\phi=0)$ changes by tuning $V_\mathrm{bg}$. Two methods, phase coherent and phase incoherent, are used to calculate $T(\phi=0)$ as displayed by the red and black lines, respectively. }
		\label{S_BL12C_FP_Gmap}}
\end{figure}

Here, we show more results for the quasiparticle tunneling in gapped BLG.
First of all, we present the characterization of device PNJ-B at $4.2\unit{K}$ and $B=0$, as the conductance map shown in Fig.~\ref{S_BL12C_FP_Gmap}(a). 
The corresponding simulation result displayed in Fig.~\ref{S_BL12C_FP_Gmap}(b), corroborates the experimental observation of FP interferences, which is quite similar to that for device PNJ-A.
When we examine the oscillatory part of the conductance $G_\mathrm{osc}$ in the $n\bar{p}nn$ regime at $B=0$, the phase shifts are observed in both experiments and simulations, as highlighted by the green dashed lines in Fig.~\ref{S_BL12C_FP_Gmap}(c) and (d), respectively.
These phase shifts are due to the sudden Berry-phase change as discussed in the main text.
Comparing Fig.~\ref{S_BL12C_FP_Gmap}(c) and Fig. 4(a) of the main text, we found that the positions where the phase shifts occur, are different. 
This discrepancy requires further understanding. 
The Berry phases~\cite{berry1984quantal,xiao2010berry} for regions T (\smash{$\Phi^\mathrm{\!\mathsmaller{ (T)}}_\mathrm{\!\mathsmaller{Berry}}$}) and B (\smash{$\Phi^\mathrm{\!\mathsmaller{ (B)}}_\mathrm{\!\mathsmaller{Berry}}$}) are presented as a function of $V_\mathrm{tg}$ and $V_\mathrm{bg}$ in Fig.~\ref{S_BL12C_FP_Gmap}(e) and (f), respectively. 
The Berry phase for T changes with respect to both $V_\mathrm{tg}$ and $V_\mathrm{bg}$, while the Berry phase for B only depends on $V_\mathrm{bg}$.

We show a simulation example of FP interferences under low magnetic fields in Fig.~\ref{S_BL12C_FP_Gmap}(g), where the single-mode conductance $g$ varies as a function of $V_\mathrm{bg}$ and $B$ at $V_\mathrm{tg}=-1.5$ V. 
The conductance oscillations exist at $V_\mathrm{bg}$ in the range of $-7.5\sim30\unit{V}$ , which is in the $n\bar{p}nn$ regime.
Instead of tuning $V_\mathrm{tg}$ as in Fig.~3(a) of the main text, we detect the SLG-like (SLG for single layer graphene) phase shifts of the fringes at $B\approx100$--$200\unit{mT}$ by changing $V_\mathrm{bg}$. 
The black solid and red dashed lines in Fig.~\ref{S_BL12C_FP_Gmap}(g) show the initial and shifted positions of the FP fringes, respectively.
The half-period shift occurs at $V_\mathrm{bg}\approx22\unit{V}$, indicating the Berry phase of $\pi$ is picked up in region T.
The corresponding transmission probability at normal incidence $T(\phi=0)$ is calculated for two cases, phase coherent and phase incoherent, shown as the red and black curves in Fig.~\ref{S_BL12C_FP_Gmap}(h), respectively.
Here, we found the transition from anti-Klein to nearly complete Klein tunneling as in the main text when $V_\mathrm{bg}$ declines from $29\unit{V}$ to $10\unit{V}$.
The maximum of $T(\phi=0)$ is 0.86 at $V_\mathrm{bg}=22\unit{V}$.
For $V_\mathrm{bg}<22\unit{V}$, $T(\phi=0)$ drops to 0.32, in contrast to 0.65 obtained in Fig.~3(i) in the main text.
Therefore, the recovery of anti-Klein tunneling at high Fermi energies is improved, since the Berry phases in B and T are both closer to 0 (or equivalently $2\pi$).


%